\documentclass[a4paper]{aa}
\usepackage{txfonts}
\usepackage{natbib}
\usepackage{graphicx}

\bibstyle{aa}

\newcommand{\fifps}[2]{\centering\resizebox{#1}{!}{\includegraphics{#2}}}

\newcommand{\filps}[2]{\resizebox{#1}{!}{\rotatebox{-90}{\includegraphics{#2}}}}

\def\llm{{\sc LLModels}}

\def\atlas{{\sc ATLAS9}}
\def\tef{T_{\rm eff}}
\def\tauros{\tau_{\rm ross}}
\def\taucont{\tau_{5000}}

\def\paperone{{Paper~I}}
\def\papertwo{{Paper~II}}

\begin{document}

\title{Stellar model atmospheres with magnetic line blanketing}
\subtitle{III. The role of magnetic field inclination}
\titlerunning{Stellar model atmospheres with magnetic line blanketing. III.}

\author{S. A. Khan\inst{1,2} \and D. V. Shulyak\inst{2}}

\offprints{S. A. Khan, \email{skhan@astro.uwo.ca}}

\institute{%
Physics and Astronomy Department, University of Western Ontario, London, ON, N6A 3K7, Canada \and
Institut f\"ur Astronomie, Universit\"at Wien, T\"urkenschanzstra{\ss}e 17, 1180 Wien, Austria }

\date{Received 31 January 2006 / Accepted 29 March 2006}
\abstract
{We introduced the technique of model atmosphere calculation with polarized radiative transfer and magnetic line blanketing. However, the calculation of model atmospheres with realistic magnetic field configurations (field strength and angle defined relative to the atmosphere plane) has not been previously attempted.}
{In the last paper of this series we study the effects of the magnetic field, varying its strength and orientation, on the model atmosphere structure, the energy distribution, photometric colors and the hydrogen Balmer line profiles. We compare with the previous results for an isotropic case in order to understand whether there is a clear relation between the value of the magnetic field angle and model changes, and to study how important the additional orientational information is. Also, we examine the probable explanation of the visual flux depressions of the magnetic chemically peculiar stars in the context of this work.}
{We calculated one more grid of the model atmospheres of magnetic A and B stars for different effective temperatures (${\tef=8000}$\,K, 11\,000\,K, 15\,000\,K), magnetic field strengths (${B=0}$, 5, 10, 40\,kG) and various angles of the magnetic field (${\Omega=0\degr\!}$--$\,90\degr$) with respect to the atmosphere plane. We used the \llm\ code which implements a direct method for line opacity calculation, anomalous Zeeman splitting of spectral lines, and polarized radiation transfer.}
{We have not found significant changes in model atmosphere structure, photometric and spectroscopic observables or profiles of hydrogen Balmer lines as we vary the magnetic field inclination angle $\Omega$. The strength of the magnetic field plays the main role in magnetic line blanketing. We show that the magnetic field has a clear relation to the visual flux depressions of the magnetic CP stars.}
{We can use the approach introduced in the previous paper of this series, which neglects anisotropy effects, to calculate model atmospheres with magnetic line blanketing. This technique seems to be reliable, at least for homogeneous atmospheres with scaled solar abundances.}

\maketitle

\section{Introduction}
In the previous papers of this series (Khan et al. 2004; Kochukhov et al. 2005, \paperone; Khan \& Shulyak 2006, \papertwo) we studied the effects of magnetic line blanketing on model atmospheres of magnetic chemically peculiar (CP) stars. We showed that models with magnetic line blanketing can produce the characteristic observable features which are typical of the magnetic CP stars, such as flux redistribution from the UV region to the visual \citep{leckrone1} and flux depression around 5200\,\AA\ \citep{kodaira}. In our earlier work we made some assumptions that simplified the modelling technique but still allowed us to carry out a systematic analysis of the role of the magnetic field in the line blanketing and its influence on the observables.

The magnetic field influences the formation of spectral lines due to the anomalous Zeeman effect. Spectral lines split into a number of components, increasing the line absorption and the line equivalent width, which is known as the magnetic intensification effect. Hundreds of thousands of lines modify the line opacity, and then change the model atmosphere structure (due to the backwarming effect), as well as modifying the outgoing radiation. The specific line profile in the presence of the magnetic field depends on the anomalous Zeeman splitting pattern, the magnetic field strength and the inclination angle between the magnetic field vector and the light propagation direction (often named the line of sight). The strength of the magnetic field and the splitting pattern determine the wavelength shift of the split components, while the shape of component profiles is defined by the inclination angle and the splitting pattern via solution of the radiative transfer equation for polarized light.

In a magnetic plane-parallel stellar atmosphere, the light propagating in different directions has different angles to the magnetic field vector, and thus carries different amounts of energy. However, one can expect that the overall effect of the enormous number of split spectral lines on line blanketing depends mainly on number of these components and the magnetic field magnitude. The point is that for different patterns the maximum of the magnetic intensification occurs for different inclination angles \citep[][Sect.~3.1]{stift}, and particular angle dependent variations of split line profiles may appear to be insignificant on the line blanketing after averaging through the solid angle integration procedure for millions of components. Consequently, the approach which neglects the dependance of the inclination angle of the magnetic field over different pencils of radiation and adopts the same angle for all of them looks adequate. Such an approach was used in \paperone\ and \papertwo.

Looking back, we can say that \paperone\ revealed the general effects of the enhanced line blanketing due to Zeeman splitting of spectral lines on the backwarming effect, and resulting effects in producing signatures on the outgoing radiation. In \papertwo\ we introduced polarized radiation transfer in the magnetic model atmospheres and developed the technique of such model atmosphere calculations. We showed that neglecting of effects of polarized radiative transfer overestimates the influence of the Zeeman splitting on the magnetic opacities, and thus can not be used as an accurate approximation.

One aspect that remains to be included for absolutely correct magnetic line blanketing modelling is to take into account variation of the inclination angle over different directions in the stellar atmosphere. This will allow us to conduct a comprehensive analysis of the approximation adopted in \papertwo\, to find probable systematic errors such as over- or underestimation of the line blanketing effect.

In this final paper of the series we compute a grid of LTE 1D model atmospheres of A and B stars with the same fundamental parameters used in previous works but replacing the simplified approach in which the magnetic field is assigned with respect to the light propagation direction. Here we specify exactly the strength and the angle of the magnetic field with respect to the atmosphere plane, and vary them across the fundamental modelling parameters space. We then study the resulting effects on the magnetic line blanketing and the consequent changes in the model atmosphere structure and the emergent flux.

The paper is organized as follows. In Sect.\,\ref{techniques} we describe new aspects of the magnetic model atmosphere calculation. Sect.\,\ref{results} presents numerical results and extensive analysis of the flux distribution and Sect.\,\ref{discussion} summarizes results of this work.

\section{Calculation of magnetic model atmospheres}
\label{techniques}
\begin{figure}
\fifps{5cm}{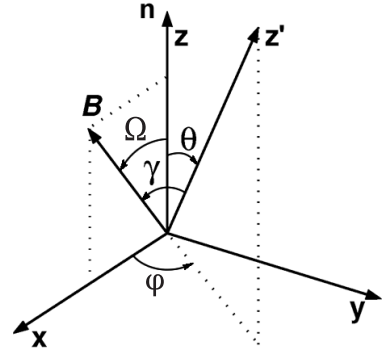}
\caption{Reference frame of plane-parallel atmosphere. The $z$-axis is parallel to the unit vector $\vec{n}$ and the $xy$ plane coincides with the plane of an atmosphere. The $z'$-axis (propagation direction of light) has an inclination $\theta$ to the unit vector $\vec{n}$ and azimuth $\varphi$ measured counterclockwise from the $x$-axis: ${0\leq\theta\leq\pi}$\,,\,\,\,${0\leq\varphi\leq 2\pi}$. The magnetic vector $\vec{B}$ is located in the $xz$ plane, has an inclination $\gamma$ to the propagation direction and an inclination $\Omega$ (${0\leq\Omega\leq\pi}$) to the unit vector. The $x'$ and $y'$ axes are not shown in figure because their choice does not modify the main equations used in this paper.}
\label{reference2}
\end{figure}

The technique of model atmospheres calculation with polarized radiative transfer and magnetic line blanketing due to the  anomalous Zeeman splitting is presented in \papertwo\ in detail. In this work we fully implement this technique, dropping an assumption about isotropic properties of the stellar atmosphere penetrated by the magnetic field, which was adopted in \paperone\ and \papertwo. Model atmospheres calculated in those papers were characterized only by the magnetic field strength, and assumed the same angle ${\gamma=\pi/2}$ between the pencil of radiation and the magnetic field vector, for every light propagation direction. The models computed in this paper are explicitly defined not only by the strength of the magnetic field but also by its inclination angle $\Omega$ to the vertical (see \papertwo, Sect.~2.5). Thus the angle $\gamma$ depends on the radiation direction, introducing the anisotropy to the modelling, and we have to modify the integration scheme over solid angle in order to deal with this. For further explanation see Fig.~\ref{reference2} (which is the same as Fig.~2 in \papertwo).

In the general case, the integration procedure is performed over inclination angle $\theta$ to the vertical (${0\leq\theta\leq\pi}$) and azimuthal angle $\varphi$ (${0\leq\varphi\leq 2\pi}$). In a case of an isotropic medium (the intensity does not depend on azimuth) the integration is performed over ${\mu=\cos\theta}$ within the limits ${0\leq\mu\leq 1}$ instead of ${-1\leq\mu\leq 1}$, splitting the integral into two for outward and inward intensities.

For model atmospheres with magnetic line blanketing it is also possible to split the integral and reduce the coverage of $\mu$ to the interval ${0\leq\mu\leq 1}$, accompanied by the reducing of the interval for the $\varphi$ angle to limits ${\,0\leq\varphi\leq\pi}$. This follows from several properties of the symmetry in the magnetic case. The first one is that for inward radiation the rotation of the model atmosphere reference frame $xyz$ about the $x$-axis by ${\pi}$ (the inclination angle $\gamma$ of the magnetic field to the pencil of radiation converts to the angle ${\pi-\gamma}$) does not change the total intensity $I$ of the beam. The second and third ones arise from the expression for $\cos\gamma$ (see Eq.~(46) in \papertwo)
\begin{equation}
\cos\gamma=\cos\theta\cos\Omega+\sin\theta\sin\Omega\cos\varphi,
\end{equation}
which is illustrated in Fig.~\ref{angles}. One can see the mirror symmetry for $\cos\gamma$ with respect to the line ${\varphi=\pi}$ and the symmetry of rotation around the point ($\pi/2$; 0) for ${\,0\leq\varphi\leq\pi\,}$. In other words, the $xz$ plane which contains the magnetic vector $\vec{B}$ naturally divides both the top (outward radiation) and the bottom (inward radiation) integration domains (hemispheres) into two equal integrals. At the same time, similar to the isotropic case, the integration over the inclination $\theta$ is also reduced by half. In fact, the change of the angle $\gamma$ to ${\pi-\gamma}$ and general rotation of the local propagation reference frame ($x'y'z'$) about the $z'$-axis do not change the flux or radiative equilibrium equations, so Eqs.~(42) and (44) in \papertwo\ hold directly in this study.

Consequently, the integration over a solid angle is carried out on a 2D ($\theta$, $\varphi$) grid for ${\,0\leq\theta\leq\pi/2}$ and ${\,0\leq\varphi\leq\pi}$ that corresponds to a quarter of a sphere. The integration is performed over 4 quadrature points in ${\mu=\cos\theta}$, and over 8 points in azimuthal angle $\varphi$, since the range of $\varphi$ is twice large as that of $\theta$; in total we use 32 points. The numerical integration employs Gauss-Legandre quadratures. Our numerical tests show that this number of points is absolutely sufficient for the integration procedure. Even an integration using 3 by 6 points provides an accuracy within 1\%. This conclusion coincides with the result of extensive tests by \citet[][Sect.~5.1]{stift_rad1}.

\section{Numerical results}
\label{results}
\begin{figure}
\fifps{\hsize}{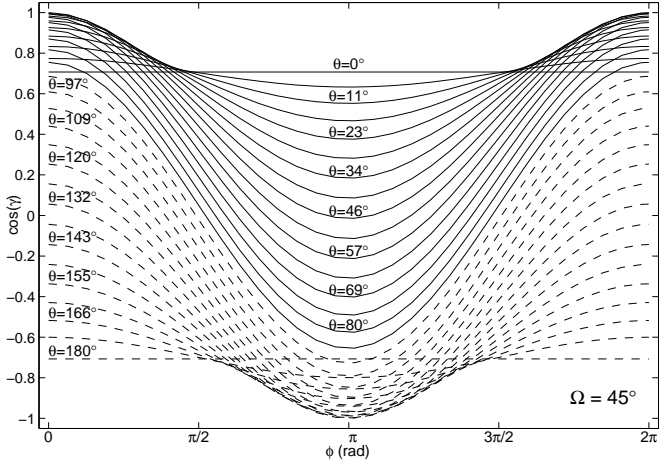}
\caption{The dependance of $\cos\gamma$ on the azimuthal angle $\varphi$ for different inclination angles $\theta$. The angle $\Omega$ of the inclination of the magnetic field is set to 45$\degr$. The figure illustrates the symmetric properties of the $\cos\gamma$ values. The solid lines represent curves for ${\,0\leq\theta\leq\pi/2}$\,, the dashed lines for ${\,\pi/2<\theta\leq\pi}$.}
\label{angles}
\end{figure}

In order to be consistent with previous studies we have used the same fundamental parameters as in \papertwo. The stellar effective temperature assumed has the values ${\tef=8000}$\,K, 11\,000\,K, 15\,000\,K, the surface gravity ${\log g=4.0}$, metallicities ${[M/H]=0.0}$, +1.0 (for non-solar metallicities, the He abundance was decreased to normalize the sum of all abundances), magnetic field strengths 0, 5, 10, and 40\,kG, with inclination $\Omega$ to the atmosphere plane ${\Omega=0\degr}$, 45$\degr$ and 90$\degr$ (for two subsets of models the angle $\Omega$ ranges from $0\degr$ to $90\degr$ in $15\degr$ steps). The parameter values that we have assumed approximately cover the stellar parameter space occupied by the magnetic Ap and Bp stars.

We used the logarithmic continuum optical depth scale $\log\taucont$ running from $+2$ to $-6.875$ and subdivided into 72 equally-spaced layers as a scale for the model atmospheres calculation. Convection and microturbulent velocity were neglected because of the presence of a global magnetic field and direct inclusion of the magnetic intensification. The wavelength ranges for the spectrum synthesis are as specified in \papertwo\ (Sect.~3).

For model atmosphere calculation we used a line list extracted from the VALD database \citep{vald} including lines originating from predicted energy levels. The procedure of the preparation of the line list and dealing with Land\'e factors is described in \papertwo\ (Sect.~3.1). Here we recall that Zeeman splitting is taken into account for all spectral lines except the hydrogen lines.

We then used this line list for line preselection in the \llm\ code, using the selection threshold 1\% (see \papertwo, Sect.~3.1). For each set of $\tef$, $[M/H]$, $B$ values, we have used a corresponding converged model calculated in \papertwo\ as an initial guess of the model atmosphere structure and for the spectral line preselection procedure.

In the following sections we study the model temperature structure, spectral energy distribution, photometric colors and profiles of hydrogen Balmer lines. Our main interest is to compare them to results obtained in the previous paper, i.e. to model atmospheres calculated assuming that the orientation of the magnetic field is characterized by the same angle ${\gamma=\pi/2}$ for every light propagation direction. In those models we neglect the anisotropy induced by the magnetic field, because the magnetic field is specified with respect to the local directions of the pencil of radiation, not to the atmosphere plane. For future reference we will call these model atmospheres ``reference magnetically isotropic models".

To perform a self-consistent study and eliminate possible systematic errors, we recalculated reference magnetically isotropic models because a new flux integration scheme was introduced and a new version of the \llm\ code was issued. We also reduced the number of model atmospheres to be computed in comparison to our usual grid: we did not consider models with intermediate metallicity value ${[M/H]=+0.5}$ and magnetic field strengths ${B=1}$\,kG and 20\,kG. This allowed us to save computational time without losing a view of the overall picture.

We note that the new models are pretty time consuming; one iteration takes about 7 hour on a modern PC or laptop, in comparison to about 2 hours and 50 minutes for models calculated in \papertwo\ and \paperone\ respectively, while one iteration for non-magnetic models takes about 5 minutes. To get converged the models presented in this paper requires 6--12 iterations. In \paperone\ and \papertwo\ the number of iterations was 7--12 and 9--20 respectively, depending on modelling parameters. We recall that in this series we used simplified or previously calculated model atmospheres (with the same $\tef$, $\log g$ and $[M/H]$ or even with the same magnetic field strength) as an initial guess of the model atmosphere structure. This provides us a very good starting point to iterate a new model with the more sophisticated approach. Usually, for new models, five iterations are enough to achieve convergence throughout the whole atmosphere except surface layers where the rest of the iterations are required to correct the temperature of these layers by just 2--10\,K. Poor convergence in surface layers is a known issue since \atlas, and controlled by the condition of radiative equilibrium (\papertwo, Sect.~2.3). Essentially, direct consideration of the anomalous Zeeman splitting increases the number of spectral lines ($\pi$- and $\sigma$-components) involved in calculations by a factor of about 17. Moreover, the integration is carried out on a 32 point 2D grid instead of the usual 3 $\mu$-point 1D grid, and the solution of the polarized radiative transfer equations takes more time than the unpolarized one. The computation time does not grow linearly with the number of integration points since a lot of basic physics involved into calculations (like the very time consuming sums of Voigt and Faraday-Voigt profiles, see Eqs.~(17) and (18) in \papertwo) needs to be computed only once for all integration points at a given wavelength.

With the same goal of saving computational time, we limited ourselves to three values of the inclination angle ${\Omega=0\degr}$, 45$\degr$ and 90$\degr$ for each model of the grid. However, before doing this we checked for two different $\tef$ and $B$ values that the variation of the $\Omega$ angle on a fine angle grid produces smooth and consecutive changes on the model atmosphere structure, and ${\Omega=0\degr}$ and 90$\degr$ correspond to boundaries of these changes.

An important point about model atmospheres with magnetic line blanketing that we must discuss to avoid confusion, is determination of the stellar flux. The flux that comes from a star to the observer from the sum of all visible stellar surface unit areas, we call the stellar flux. This is exactly the flux that observations provide us as energy distribution or photometric colors. In order to compare theoretical predictions with observed values we need to compute the stellar flux.

In a isotropic medium (absorption coefficient does not depend on direction) such as a non-magnetic plane-parallel atmosphere, the intensity of the radiation depends only on its angle ${\mu=\cos\theta}$ to the vertical. At the same time, for uniform spherical stars, the variation of the intensity across the visible stellar disk has the same dependence as the local angular variation, and we can cover the stellar surface with the identical models. Thus, the calculation of the surface flux per unit area in a plane-parallel atmosphere automatically produces the stellar flux, which can be compared with observations.

In an anisotropic medium, such as a medium penetrated by a magnetic field, the extinction depends on the light propagation direction, thus the flux from a point on the atmosphere surface does not necessarily equal the stellar flux. In the general case of a non-uniform stellar surface, the star has to be covered by a grid of different model atmospheres, i.e. as applied to our case models need to be calculated with individual values of the strength and inclination of the magnetic field. Then, calculating the intensity of radiation coming from each local model to the observer, and performing the integration over the visible spherical surface in the observer reference frame, we obtain the stellar flux.

Since the real configuration of a global stellar magnetic field is pretty complicated, and its strength can vary substantially over the stellar surface, it was difficult to restrict our study to certain magnetic field configurations. In this paper our focus is to study the magnetic line blanketing effect as a function of the strength and inclination of the magnetic field in a plane-parallel atmosphere. The modelling of complicated configurations of the magnetic field naturally would bring us to modelling of particular objects, accompanying by much computational expense, and would not illuminate clearly the basic effects of magnetic line blanketing on the outgoing stellar radiation.

Thus, we decided to consider the surface flux produced by the atmosphere model as a stellar flux. This approach corresponds to the case in which the magnetic field has the same strength over the stellar surface, and the angle between the observer's line of sight and the local magnetic field vector varies in the same way over the stellar surface as the angle $\gamma$ between a local pencil of radiation and the vector of the magnetic field in a plane-parallel atmosphere. For example, for ${\Omega=0\degr}$, i.e. for a magnetic field that is perpendicular to the atmosphere plane, the surface flux produced by the model atmosphere corresponds to the radial magnetic field configuration in which the magnetic field vector is perpendicular to the stellar surface at each point. As another example, for ${\Omega=90\degr}$, the case in which the magnetic field vector lies in the atmosphere plane, the surface flux produced by the model atmosphere corresponds to a case in which the magnetic field vector is a tangent to the stellar surface at each point. Obviously, that two these cases are quite different and should give us an idea what kind of changes we may expect in energy distribution. For other $\Omega$ angle values between $0\degr$ and $90\degr$ it is rather difficult to draw a mental picture of the stellar magnetic field configuration. Some ideas of possible fields can be obtained from Fig.~\ref{angles}.

Note that considerations about the stellar flux outlined above have nothing to do with the model atmosphere calculation technique or model atmosphere structure, and concern only interpretation of the spectroscopic observables.

\subsection{Model structure}
\label{structure}
\begin{figure*}
\filps{\hsize}{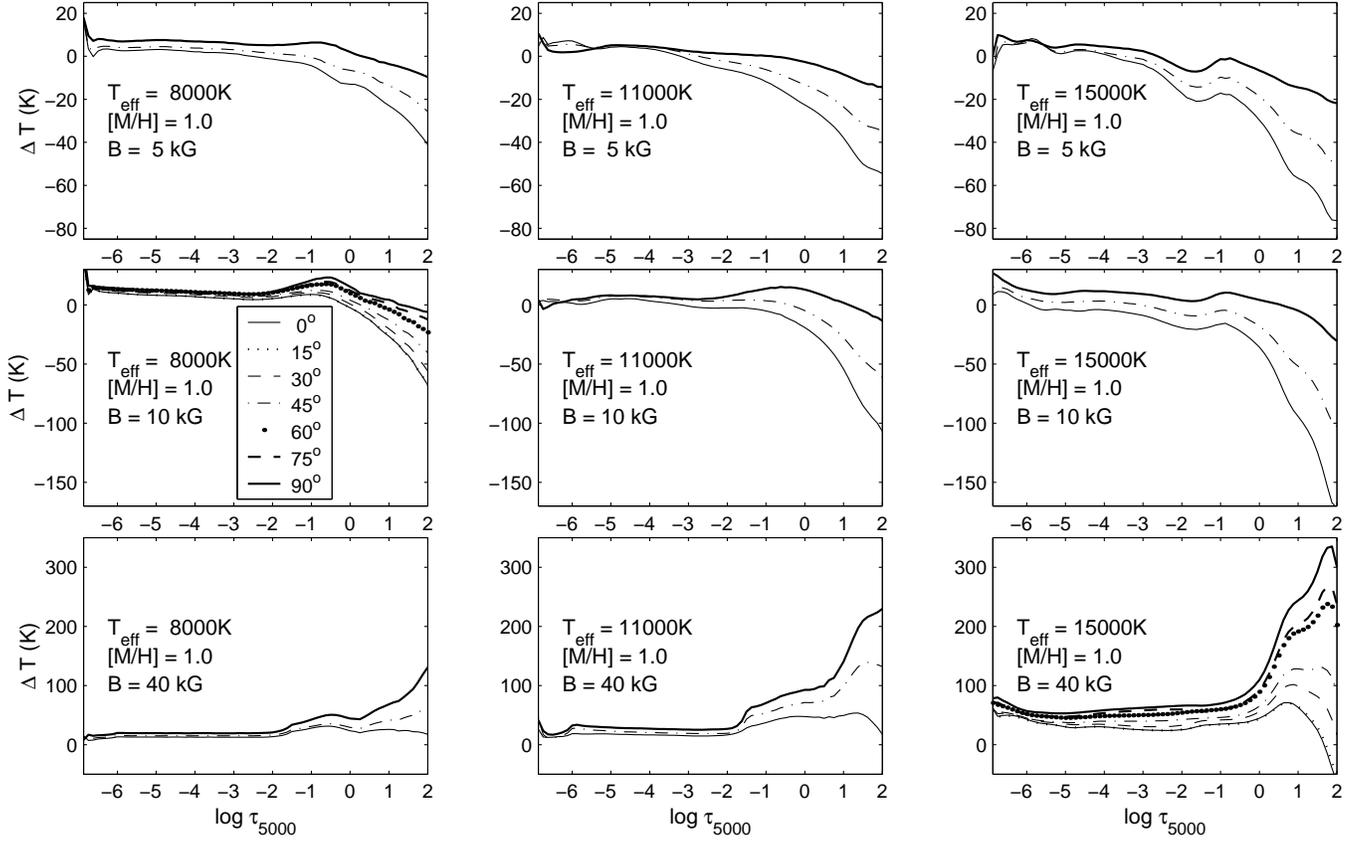}
\caption{The temperature differences between magnetic angle dependent and reference magnetically isotropic model atmospheres for the effective temperatures ${\tef=8000}$\,K, 11\,000\,K, 15\,000\,K, metallicity ${[M/H]=+1.0}$, strengths of the magnetic field ${B=5}$, 10, 40\,kG, and angles of the magnetic field $0\degr$, $45\degr$ and $90\degr$. Two frames for ${\tef=8000}$\,K, ${B=10}$\,kG and ${\tef=15\,000}$\,K, ${B=40}$\,kG represent the same dependence for a fine set of angles ranging from $0\degr$ to $90\degr$ in $15\degr$ steps. The depth scale $\log\taucont$ is a logarithmic continuum monochromatic optical depth scale calculated at $5000$\,\AA.}
\label{temperature}
\end{figure*}

In Fig.~\ref{temperature} we show the difference between the temperature structure of magnetic angle dependent, and reference magnetically isotropic models, as a function of the continuum optical depth $\log\taucont$ for metallicity ${[M/H]=+1.0}$.

We note that it is difficult to analyse the temperature changes using the $\log\taucont$ scale, in view of the backwarming effect, since the monochromatic continuum depth scale $\taucont$ does not contain frequency averaged mean absorption coefficient (unlike the Rosseland $\tauros$ scale). But having model atmospheres computed in $\log\tauros$ scale (\paperone, Fig.~1) and replotting the temperature distributions for them on the $\log\taucont$ scale, we can say that the temperature trends are pretty similar in both scales from the top of the atmosphere down approximately to ${\log\tauros=\log\taucont=-2}$. This comparison will help us shortly.

Though, in general, effects on the model structure in this work are small, two interesting features can be pointed out. The first one is that some essential properties in the model atmosphere structure appear only for depths deeper that ${\log\taucont=-1}$, while the rest of the atmosphere upward from this depth is almost unchanged. Taking into account the fact about the similarity between changes of the temperature distribution in $\log\taucont$ and $\log\tauros$ scales for the optically thin upper region, we suppose that the variation of the inclination angle $\Omega$ does not have a substantial influence on the line blanketing (and thus on the backwarming effect), since there are not any characteristic distinctions between temperature distributions for different $\Omega$ angles in upper layers (Fig.~\ref{temperature}).

The second point is that the model structure in the deeper layers demonstrates a clear dependency on the variation of the inclination angle $\Omega$ of the magnetic field. The larger the angle is, the hotter the atmosphere is. The fine angle grid subsets of models for ${\tef=8000}$\,K, ${B=10}$\,kG and ${\tef=15\,000}$\,K, ${B=40}$\,kG show this orderly behaviour in detail in Fig.~\ref{temperature}. Note, that the model structures for ${\Omega=0\degr}$ and $15\degr$ are absolutely indistinguishable. That is an interesting point since increasing of the $\Omega$ angle beyond the $15\degr$ value produces clear changes in the model atmosphere structure. It looks like this behaviour is connected with the dependance of the magnetic intensification on the field inclination angle.

The solar metallicity (${[M/H]=0.0}$) models show the same features and temperature behaviour as discussed above but the effect in model structure is approximately four times smaller than for overabundant (${[M/H]=+1.0}$) models, and is thus completely negligible.

\subsection{Energy distribution}
\label{energy}
We performed a comparison of the energy distributions of models with different $\Omega$ angles for each combination of the considered modelling parameters $\tef$, ${[M/H]}$, and $B$, relative to the reference magnetically isotropic models. We look particularly for changes in features which were found to be sensitive to the magnetic field in the previous papers: the flux redistribution from the ultraviolet region to the visual, and flux depression around 5200\,\AA.

We found that the model atmospheres calculated with different value of the angle $\Omega$ produce fluxes that are very close each to other and to the reference model flux. The discrepancy in the UV region is slightly more recognizable than in the visual and can reach as much as only 0.04\,mag (on the $\log F_\lambda$ scale) at some UV wavelength points, but no common trends were found. For a strong magnetic field ${B=40}$\,kG and overabundant composition ${[M/H]=+1.0}$, the reference magnetically isotropic models with effective temperatures ${\tef=8000}$\,K and 11\,000\,K produce UV flux that is a little bit less deficient than angle dependant models with the same fundamental parameters, so a small offset appears. The typical value of this offset amounts approximately to 0.06\,mag for ${\tef=8000}$\,K and to 0.05\,mag for ${\tef=11\,000}$\,K.

\subsection{Colors}
\label{colors}
In order to perform a complementary quantitative analysis of possible flux changes in the visual region we calculated photometric colors using energy distributions. In this work we consider the Str\"omgren-Crawford $uvby{\rm H}\beta$ system, the peculiarity index $a$, and the peculiarity indices ${V_1-G}$ and $Z$ which are linear combinations of the Geneva photometric indices. The meanings of these indices, corresponding references and a description of a computational procedure can be found in \paperone\ (Sect.~3.3).

We note that we have not shown any dependency of the photometric indices with variation of the angle $\Omega$, primarily because such variations are rather small. For the same reason a graphical representation of color values is not illuminating, and we provide a short description only.

For color analysis we consider two difference values defined among models with different angle $\Omega$ while the rest of modelling parameters remain intact. The first one we call deviation; it equals the difference between maximal and minimal index values. The second one we call the offset; it equals the maximum color change with respect to the reference magnetically isotropic model. Clearly, the absolute value of the maximum offset includes a deviation value.

We found that the deviation does not exceed 1\,mmag for peculiar indices ($a$, $Z$ and ${V_1-G}$), and 3\,mmag for all indices of the $uvby{\rm H}\beta$ Str\"omgren-Crawford system.

For the extreme case of a strong magnetic field (${B=40}$\,kG) and overabundant chemical composition ${[M/H]=+1.0}$ the deviations above are increased by 1\,mmag for the $a$ and $m_1$ indices for ${\tef=8000}$\,K, and for the $c_1$ index for ${\tef=11\,000}$\,K.

The color offset for all indices and all calculated models is usually not more than $\pm3$\,mmag, with no noticeable trends, so it is very difficult to say that there is any systematic differences in comparison to the reference model colors, although some exceptions appear.

For the metallicity indicator index $m_1$ for models with ${\tef=8000}$\,K and ${[M/H]=+1.0}$ the offset grows from $+5$\,mmag for ${B=5}$\,kG up to +20\,mmag for ${B=40}$\,kG. Another exception appears for the peculiar $a$ index for ${\tef=8000}$\,K, ${B=40}$\,kG model: the offset of the $a$ index for ${[M/H]=0.0}$ is $+5$\,mmag and for ${[M/H]=+1.0}$ is $+13$\,mmag. For Z and $c_1$ indices exceptions exist for ${[M/H]=+1.0}$, ${B=40}$\,kG models: the Z index shows the offset $-6$\,mmag for ${\tef=8000}$\,K, and the offset of the $c_1$ index reaches value $-7$\,mmag for ${\tef=15\,000}$\,K.

It is clearly seen that almost all of the exceptions occur only for the extremely strong magnetic field value 40\,kG and overabundant chemical composition ${[M/H]=+1.0}$.

\subsection{Hydrogen line profiles}
\label{hydrogen}
Since hydrogen lines are often used for determination of the stellar fundamental parameters we checked the influence of changes in the model atmosphere structure on the hydrogen Balmer lines. We calculated profiles of H$\alpha$, H$\beta$ and H$\gamma$ and compared them to the corresponding profiles computed for the reference magnetically isotropic models.

We found that the variation of the angle $\Omega$ produces differences in hydrogen profiles that are almost negligible with respect each to other, as well as with respect to profiles of magnetically isotropic models. The common deviation value for the overwhelming majority of models is less than 0.1--0.2\% of the continuum level and reaches the maximum level of 0.4--0.5\% for ${\tef=11000}$\,K, ${[M/H]=+1.0}$, ${B=40}$\,kG model for the H$\beta$ and H$\gamma$ lines.

We note that hydrogen line profiles calculated within the framework of this work for non-magnetic models (for a cross check between papers as our code evolves) differ from those calculated in \papertwo\ by less than by 0.1\%.

\subsection{Stellar atmospheric parameters}
\label{stellar_parameters}

\begin{figure*}[t*]
\filps{\hsize}{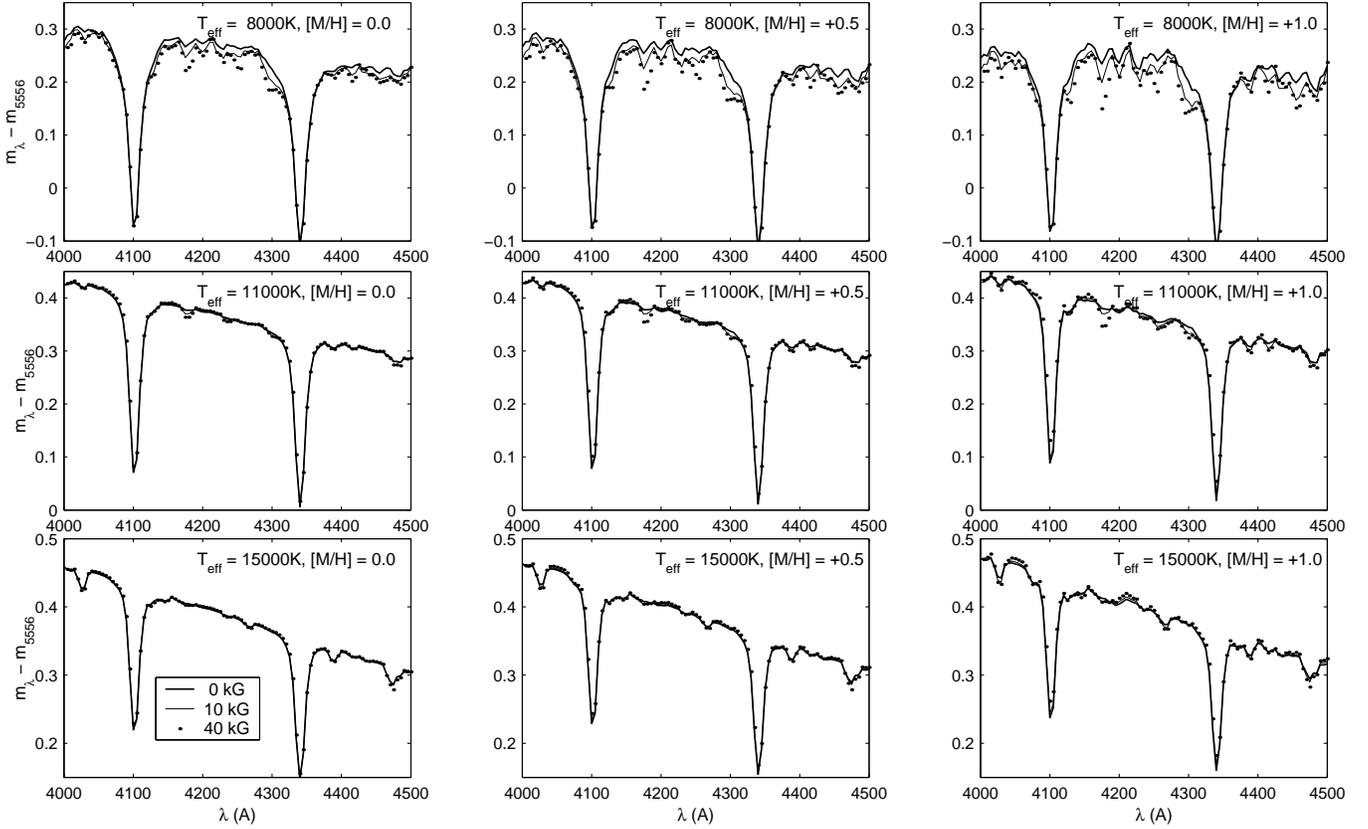}
\caption{Synthetic fluxes in the 4200\,\AA\ region normalized at 5556\,\AA\ for the effective temperatures ${\tef=8000}$\,K, 11\,000\,K, 15\,000\,K, metallicities ${[M/H]=0.0}$, $+0.5$, $+1.0$, and the magnetic field strengths ${B=0}$, 10, 40\,kG. Original stellar fluxes are convolved with a Gaussian profile with ${{\rm FWHM}=15}$\,\AA.}
\label{4200}
\end{figure*}

\begin{figure*}[t*]
\filps{\hsize}{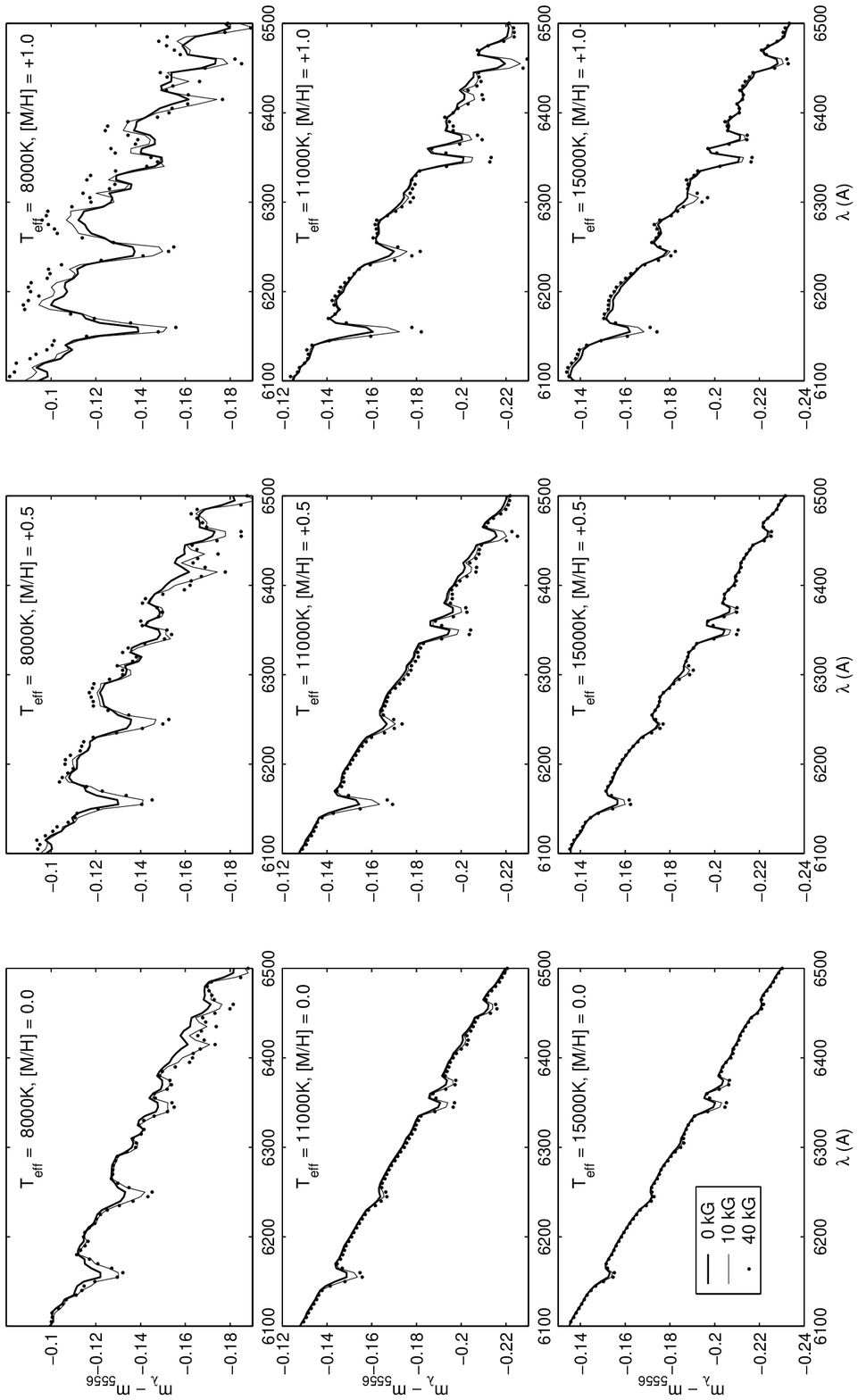}
\caption{Synthetic fluxes in the 6300\,\AA\ region normalized at 5556\,\AA\ for the effective temperatures ${\tef=8000}$\,K, 11\,000\,K, 15\,000\,K, metallicities ${[M/H]=0.0}$, $+0.5$, $+1.0$, and the magnetic field strengths ${B=0}$, 10, 40\,kG. Original stellar fluxes are convolved with a Gaussian profile with ${{\rm FWHM}=15}$\,\AA.}
\label{6300}
\end{figure*}

Since the overall modification of the photometric indices of the $uvby{\rm H}\beta$ system found in this work (Sect.~\ref{colors}) is quite small we do not consider the influence of the angle $\Omega$ variation on the photometric determination of the stellar atmospheric parameters.

However, the insensitivity of the photometric indices to the inclination of the magnetic field allows us to reconsider results found in \papertwo\ with more confidence. In the previous papers we found that the magnetic opacity does not produce systematic errors in the photometric determination of the atmosphere parameters of CP stars. The $\tef$ and $\log g$ values of the magnetic models derived from the photometric calibration for normal stars differ from those for non-magnetic models within the usual error bars. It appears that the flux redistribution from the UV to the visual region and flux depression features do not affect essentially the general photometric dependencies used for stellar atmosphere parameters determination. We suppose that this happens mainly because the slope of the Paschen continuum does not suffer much from the energy redistribution phenomenon. The energy excess in the whole visual region appears to be evenly shifted along the vertical axis only (\papertwo, Fig.~3).


\subsection{Visual flux depression}
\label{visual_depression}
Having the results obtained in this paper, proving that the inclination of the magnetic field is not a crucial aspect for construction of model atmospheres with magnetic line blanketing, we can now explore the fluxes of \papertwo\ with more confidence.

The spectra of magnetic CP stars show three broad band visual flux depressions around 4200\,\AA, 5200\,\AA\ and 6300\,\AA\ \citep{kodaira}. Some authors indicate slightly different central wavelengths of these features (e.g. 4100\,\AA\ and 5300\,\AA) which is not surprising because the depression around 4200\,\AA\ as well as around 6300\,\AA\ spans several hundred~\AA, while the depression around 5200\,\AA\ is almost for 1000\,\AA\ wide \citep[see][and the whole series]{adelman1,adelman19}.

The 5200\,\AA\ feature is detected in most magnetic CP stars, the 4200\,\AA\ feature does not appear for all of them, and only a small number of CP stars show the 6300\,\AA\ flux feature. The presence of all these depressions tends to be observed only for magnetic CP stars.

To study these features, the classic photometry systems such as Geneva or Str\"omgren \citep{hauck,masana} and the specialized $a$ system \citep{maitzen}, as well as spectrophotometric scans of the energy distribution, have been used.

The origin of the flux depressions are most likely connected with the line blanketing resulted from an abundance peculiarity and magnetic field effects. Recent investigations showed that model atmospheres with individual abundance patterns \citep{kupka2}, or even individual abundances combined with stratified abundances \citep{llmodels}, can explain observed energy distribution of Ap stars while models with scaled and homogeneous abundances fail to do this \citep{adelman1995}.

Thus, an individual chemical composition and a non-uniform distribution of chemical elements with depth that is observed \citep{wade} and theoretically explained \citep{leblanc_poprad} should without doubt be taken into consideration for the atmosphere modelling of individual stars.

Since this series is focused on effects of the magnetic field on the line blanketing, and does not study individual objects, we did not consider abundances other than scaled to the solar ones. This allowed us to build an uniform grid of model atmospheres to find the main dependencies between the magnetic line blanketing and changes in the energy distribution.

A part of the energy distribution analysis in this series was an investigation of the photometric indices. Here we want to call attention to the $m_1$ index. The color-index $m_1$ of the $uvby{\rm H}\beta$ system is often used as a metallicity indicator of chemically peculiar stars. Although it was not introduced as a special parameter to separate peculiar stars from normal ones, its value is determined by three narrow band filters $vby$ which are affected by the 4200\,\AA\ and 5200\,\AA\ features. As follows from Fig.~4 in \paperone, the $m_1$ index has a clear dependance on the magnetic field strength for low effective temperature. The results presented here in Sect.~\ref{colors} also suggest that $m_1$ is sensitive to the magnetic field. Thus, we decided to expand our energy distribution study and look for probable depressions around 4200\,\AA\ and 6300\,\AA\ (the flux feature around 5200\,\AA\ was well studied in \paperone\ and \papertwo).

Fig.~\ref{4200} and Fig.~\ref{6300} show the spectral energy distributions for the 4200\,\AA\ and 6300\,\AA\ regions, based on the data of \papertwo. The synthetic fluxes are convolved with a Gaussian and normalized at 5556\,\AA\ (${m_\lambda=\log F_\lambda}$). This normalization is a usual procedure for observed energy distributions, which are often normalized at 5000\,\AA\ \citep{adelman1995} or 5556\,\AA\ \citep{adelman1975}. The latter wavelength corresponds approximately to the mean wavelength 5500\,\AA\ of the Geneva $V$ photometric band. The normalization of the flux at both of these wavelengths can be undesirably affected by the nearby 5200\,\AA\ depression which decreases the true continuum level. However our purpose was to plot fluxes in a manner suitable for comparison, and we found that normalization at 5556\,\AA\ provides a better visualization without additional flux shifts along vertical axis. We note that \citet{kupka2} showed no significant differences between these two kinds of the normalization.

We found that the depression features in spectral regions around 4200\,\AA\ and 6300\,\AA\ depend on the effective temperature, metallicity and magnetic field strength. The usual dependance on the stellar effective temperature, namely that for low $\tef$ the line blanketing produces substantial changes while for higher temperatures this effect becomes rather small, takes place again. The sensitivity of depressions to the magnetic field magnitude depends on metallicity and grows with increasing metallicity. We note that the overabundant chemical composition itself for non-magnetic models has a positive effect on the depression magnitude.

The maximum deficiency produced by a magnetic field ${B=10}$\,kG compared to the non-magnetic energy distribution for the 4200\,\AA\ region amounts to 20--40\,mmag for ${\tef=8000}$\,K and 5--10\,mmag for ${\tef=11\,000}$\,K; for the 6300\,\AA\ region these values are 10--15\,mmag for ${\tef=8000}$\,K and about 5\,mmag for ${\tef=11\,000}$\,K. For the higher effective temperature of ${\tef=15\,000}$\,K, differences between magnetic and non-magnetic energy distributions almost disappear.

Since there are no well-established photometric systems like the $a$ system to measure flux depressions around 4200\,\AA\ and 6300\,\AA\ it is difficult to make a definitive conclusion about the relation to the observed values. But it is clear that the magnetic field can be an important factor in these spectral regions to increase the magnitude of the depression. Also, it should be remembered that a specific chemical composition, and especially stratified abundances, can result in more pronounced changes in the visual energy distributions than a magnetic field of moderate values, but that the magnetic field itself can easily intensify weak features and make them visible in spectra of magnetic CP stars. A comprehensive analysis of model atmospheres with individualized or stratified abundances and the resulting fluxes has never been done. While it is obvious now that the study of certain CP objects requires a specific model atmosphere \citep{kupka2,llmodels} we want to emphasize the necessity to take into account the magnetic line blanketing for a detailed study, especially since we showed that a non-zero microturbulent velocity is not able to reproduce the effects of the magnetic intensification properly (\paperone, Sect.~3.6).

\section{Conclusions and discussion}
\label{discussion}
This paper ends the series of works concerning the magnetic line blanketing in stellar atmospheres and its relation to the observable quantities of magnetic CP stars. Using the technique introduced in \papertwo, we calculated a grid of  model atmospheres of magnetic A and B stars for different strengths of the magnetic field and different angles of its inclination to the atmosphere plane. We have used this grid to analyze the behaviour of the model atmosphere structure and the observables for variety of fundamental modelling parameters, mainly to look for any discrepancy between models with different magnetic field inclination angles, and to compare these models with previous results (\papertwo).

We have not found any significant changes in model atmosphere structure, photometric indices, the energy distribution and profiles of hydrogen Balmer lines that depend on the magnetic field inclination angle $\Omega$. Moreover we found that model atmospheres for different inclination $\Omega$ angles are almost the same as models calculated neglecting the anisotropy effects (\papertwo), when the magnetic field direction is assigned with respect to the pencil of radiation, not to the atmosphere plane. Only for extreme values of the magnetic field, such as ${B=40}$\,kG, do the results indicate that magnetically isotropic models tend to produce less deficient flux redistribution and 5200\,\AA\ flux depression than magnetic angle dependent models, while there are no significant differences among these.

It appears that the magnetic field strength is the main factor to affect the line blanketing. We conclude that it is clearly possible to implement the isotropic approach (\papertwo) to calculate model atmospheres with magnetic line blanketing. This approximation looks reliable, at least for the homogeneous atmospheres with scaled to the solar abundances considered in our studies, and can decrease the computation time dramatically. For atmospheres with stratified or individualized abundances this conclusion has still to be checked. At the same time in this paper we found that variation of the scaled abundances clearly has a more pronounced effect on model atmospheres than variation of the $\Omega$ angle.

For a particular magnetic field configuration (like a combination of dipole, quadrupole and octupole) the strength of the magnetic field can vary substantially over the visible stellar surface and individual model atmospheres have to be calculated to cover it. However, taking into account a very weak dependence of the model atmosphere structure on the magnetic field inclination to the atmosphere plane found in this study the computational expenses can be decreased. The point is that there are always unit domains with the same (or almost the same) magnetic field strength on the surface integration grid. Thus, the same model atmosphere (computed only once) can be used to calculate the outward radiation intensities for a number of surface units which have very close values of the magnetic field strength but different orientations to the observer line of sight. Then, integrating individual intensities over all unit domains we obtain the stellar flux corresponding to a particular magnetic field configuration.

Also we note that in general, the magnetic fields of Ap stars are considered to be dipoles to first order. The field intensity for a dipolar field is twice as strong at the poles as at the equator. Essentially all studies using longitudinal field and surface field measurements have found that the fields deviate from a dipole -- that one pole is stronger than the other (that a quadrupolar component is required) and often that the pole-to-equator variation of the field strength is not as strong as that of a dipole (an octupolar component is required) \citep[e.g.][]{landstreet}. Consequently, a typical range of field intensity variation over the surface of Ap stars is probably about 30--40\%. Thus, we suppose that a model atmosphere calculated using the isotropic approach and the observable mean surface magnetic field strength $B_{\rm s}$ as an input model parameter seems to be reliable (say, for routine abundance analysis of magnetic Ap stars).

Furthermore, for the isotropic approach with ${\gamma=\pi/2}$ (i.e. the magnetic field is perpendicular to every pencil of radiation) and with ${\chi=0}$ (the azimuth of the magnetic field in the local reference frame $x'y'z'$, and its choice is free, see Fig.~\ref{reference2}), the line propagation matrix $\vec{\Phi}$ (Eq.\,(7) in \papertwo) is
\begin{equation}
\begin{array}{rcll}
\vec{\Phi}= & \left(
\begin{array}{llll}
\phi_I & \,\,\,\, \phi_Q & \,\,\,\, 0      & \,\,\,\, 0      \\
\phi_Q & \,\,\,\, \phi_I & \,\,\,\, 0      & \,\,\,\, 0      \\
     0 & \,\,\,\, 0      & \,\,\,\, \phi_I & \,\,\,\, \psi_Q \\
     0 & \,\,\,\, 0      &         -\psi_Q & \,\,\,\, \phi_I \\
\end{array}
\right)
\end{array}.
\end{equation}
Thus the system of four Stokes transfer equations
\begin{equation}
\mu\frac{{\rm d}\vec{I}}{{\rm d}m}=\vec{K}(\vec{I}-\vec{S}),
\end{equation}
where ${\vec{I}=(I,Q,U,V)^\dag}$ is the Stokes vector, $m$ is column mass, ${\vec{S}=(S,0,0,0)^\dag}$ is the source vector, and $\vec{K}$ is the total propagation matrix (see Eqs.\,(2--8) in \papertwo), decouples into two independent sets of two equations each one. The second set for $U$ and $V$ does not contain emission terms and is out of our interest because boundary conditions for inward and outward radiation assume initially unpolarized light (see Eqs.\,(38-39) in \papertwo). The first set for $I$ and $Q$ is
\begin{equation}
\left\{
\begin{array}{ll}
\mu\displaystyle\frac{{\rm d}I}{{\rm d}m}=\kappa_I\,(I-S)+\kappa_Q Q\,, \vspace{0.2cm} \\
\mu\displaystyle\frac{{\rm d}Q}{{\rm d}m}=\kappa_Q\,(I-S)+\kappa_I Q\,,
\end{array}
\right.
\end{equation}
where $\kappa_I$ and $\kappa_Q$ are the elements of the total propagation matrix $\vec{K}$ which correspond to $\phi_I$ and $\phi_Q$ of the line propagation matrix $\vec{\Phi}$ (see Eq.\,(3) in \papertwo).

Consequently, the transfer equation for polarized radiation reduces to the solution of two standard unpolarized radiative transfer equations with different substitutions for the intensity and absorption coefficient
\begin{equation}
\mu\frac{{\rm d}(I\pm Q)}{{\rm d}m}=(\kappa_I\pm\kappa_Q)(I\pm Q-S).
\end{equation}
The final Stokes parameters $I$ and $Q$ are defined as a half-sum and half-difference of the solutions, respectively. In that way, for the magnetically isotropic models with ${\gamma=\pi/2}$, we can save even more computational time.

We also supplemented the investigation of the most prominent feature of the CP stars in the visual region with the flux depressions centered at 4200\,\AA\ and 6300\,\AA. Our numerical results show that the magnitude of these features clearly depends on the magnetic field strength, effective temperature and metallicity. The depressions grow with increasing magnetic field strength and metallicity. However, for higher $\tef$ the magnitude and the sensitivity to the magnetic field goes down quickly, which is similar to the results for the depression around 5200\,\AA\ (\papertwo).

Taking into account the common relation between visual flux depression magnitudes around 4200\,\AA, 5200\,\AA\ and 6300\,\AA, and the effective temperature found in this series, we suppose that individual chemical composition and inhomogeneous vertical abundance distribution are most probable candidates to explain visual peculiarities of Ap and Bp stars with higher effective temperatures.

\begin{acknowledgements}
We are grateful to Prof.~J.~D.~Landstreet for his useful comments and discussions, and corrections to the final version of the manuscript. We are also thankful to Prof.~G.~A.~Wade for his helpful discussion and to the referee for his advice on the manuscript improvement. This work was supported by a Postdoctoral Fellowship to SK at UWO funded by a Natural Science and Engineering Council of Canada Discovery Grant, and by INTAS grant 03-55-652 to DS. We also acknowledge the support by the Austrian Science Fonds (FWF-P17890).
\end{acknowledgements}


\begin{thebibliography}{10}
\bibitem[Adelman, 1975]{adelman1975}Adelman, S. J. 1975, \apj, 195, 397
\bibitem[Adelman, 1979]{adelman1}Adelman, S. J. 1979, \aj, 84, 857
\bibitem[Adelman \& Pyper, 1993]{adelman19}Adelman, S. J., \& Pyper, D. M. 1993, \aaps, 101, 393
\bibitem[Adelman et al., 1995]{adelman1995}Adelman, S. J., Pyper, D. M., Lopez-Garcia, Z., \& Caliskan, H. 1995, \aap, 296, 467
\bibitem[Alecian \& Stift, 2002]{stift_rad1} Alecian, G., \& Stift, M.~J.\ 2002, \aap, 387, 271
\bibitem[Hauck, 1974]{hauck}Hauck, B. 1974, \aap, 32, 447
\bibitem[Khan et al., 2004]{khan_paper1}Khan, S. A., Kochukhov, O., \& Shulyak, D. 2004, in Proc. IAU Symp. N~224, The A-Star Puzzle (UK: Cambridge University Press), 29
\bibitem[Kochukhov et al., 2005]{paper1}Kochukhov, O., Khan, S., \& Shulyak, D. 2005, \aap, 433, 671 (Paper~I)
\bibitem[Khan \& Shulyak, 2006]{paper2}Khan, S. A., \& Shulyak, D. V. 2006, \aap, 448, 1153 (Paper~II)
\bibitem[Kodaira, 1969]{kodaira}Kodaira, K. 1969, \apj, 157, 59
\bibitem[Kupka et al., 1999]{vald}Kupka, F., Piskunov, N., Ryabchikova, T. A., Stempels, H. C., \& Weiss, W. W. 1999, \aaps, 138, 119
\bibitem[Kupka et al., 2004]{kupka2}Kupka, F., Paunzen, E., Iliev, I. Kh., \& Maitzen, H. M. 2004, \mnras, 352, 863
\bibitem[Landstreet \& Mathys, 2000]{landstreet} Landstreet, J.~D., \& Mathys, G.\ 2000, \aap, 359, 213
\bibitem[LeBlanc \& Monin, 2004]{leblanc_poprad} LeBlanc, F., \& Monin, D. 2004, in The A-Star Puzzle, ed. J. Zverko, J. \v{Z}i\v{z}\v{n}ovsk\'{y}, S. J. Adelman, \& W. W. Weiss (UK: Cambridge University Press), 193
\bibitem[Leckrone, 1973]{leckrone1}Leckrone, D. 1973, \apj, 185, 577
\bibitem[Maitzen, 1976]{maitzen}Maitzen, H. 1976, \aap, 51, 223
\bibitem[Masana et al., 1998]{masana}Masana, E., Jordi, C., Maitzen, H. M., \& Torra, J. 1998, \aaps, 128, 265
\bibitem[Shulyak et al., 2004]{llmodels}Shulyak, D., Tsymbal, V., Ryabchikova, T., St\"utz\, Ch., \& Weiss, W. 2004, \aap, 428, 993
\bibitem[Stift \& Leone, 2003]{stift}Stift, M., \& Leone, F. 2003, \aap, 398, 411
\bibitem[Wade et al., 2001]{wade}Wade, G. A., Ryabchikova, T. A., Bagnulo, S., \& Piskunov, N. 2001, in Magnetic Fields Across the Herzsprung-Russell Diagram, ed. G. Mathys, S. K. Solanki, \& D. T. Wickramasinghe, ASP Conf. Ser., 248, 373
\end{thebibliography}
\end{document}